\documentclass[useAMS,usenatbib,usegraphicx]{mn2e}

\title[The origin of fluorine in the Milky Way]
{On the origin of fluorine in the Milky Way}
\author[A.~Renda et~al.]{Agostino 
Renda$^1$\thanks{arenda,yfenner,bgibson@astro.swin.edu.au}, 
Yeshe Fenner$^1$\footnotemark[1], Brad K. Gibson$^1$\footnotemark[1],
\newauthor 
Amanda I. Karakas$^{2,3}$, John C. Lattanzio$^2$, Simon Campbell$^2$, 
Alessandro Chieffi$^{1,2,4,5}$,
\newauthor 
Katia Cunha$^6$, Verne V. Smith$^7$\vspace{0.3cm}\\
$^{1}$Centre for Astrophysics \& Supercomputing, Swinburne University, 
Hawthorn, Victoria, 3122, Australia\\
$^{2}$ Centre for Stellar \& Planetary Astrophysics, School of Mathematical 
Sciences,\\ 
Monash University, Clayton, Victoria 3800, Australia\\
$^{3}$ Institute for Computational Astrophysics, Department of Astronomy \& 
Physics, Saint Mary's University,\\ 
Halifax, Nova Scotia B3H 3C3, Canada\\
$^{4}$ Istituto di Astrofisica Spaziale e Fisica Cosmica (CNR), Via Fosso 
del Cavaliere, 00133 Roma, Italia\\ 
$^{5}$ INAF Osservatorio Astronomico di Roma, Via Frascati 33, 00040 
Monteporzio Catone, Italia\\
$^{6}$ Observat\' orio Nacional, Rua General Jos\' e Cristino 77, 
20921 400 S\~ ao Cristov\~ ao, Rio de Janeiro, Brazil\\
$^{7}$ Department of Physics, University of Texas at El Paso, 
El Paso, TX 79968, U.S.A.
}

\begin{document}

\date{Accepted. Received; in original form}

\pagerange{\pageref{firstpage}--\pageref{lastpage}} \pubyear{2004}

\maketitle

\label{firstpage}

\begin{abstract}
The main astrophysical factories of fluorine ($^{19}$F) are thought
to be Type~II supernovae, Wolf--Rayet stars, and the asymptotic giant 
branch (AGB)
of intermediate mass stars. We present a model for the chemical evolution of 
fluorine in the Milky Way using a semi-analytic multi-zone chemical evolution 
model. For the first time, we demonstrate quantitatively the impact of 
fluorine nucleosynthesis in Wolf--Rayet and AGB stars. 
The inclusion of these latter two fluorine production sites provides a possible
solution to the long-standing discrepancy between model predictions and the
fluorine abundances observed in Milky Way giants. Finally, fluorine is 
discussed as a possible probe of the role of supernovae and intermediate
mass stars in the chemical evolution history of the globular cluster
$\omega$~Centauri.
\end{abstract}
\begin{keywords}
galaxies: evolution -- stars: abundances -- stars: evolution
\end{keywords}

\section{Introduction} 

The three primary astrophysical factories for fluorine ($^{19}$F) production 
have long been thought to be 
Type~II Supernovae (SNe~II), Wolf--Rayet (WR) stars, and asymptotic giant
branch (AGB) stars (e.g. \citealt{WW}; \citealt{MA}; \citealt{ForestiniETal}; 
Mowlavi, Jorissen \&
Arnould 1998, respectively). 
Previous attempts to model the Galactic production and evolution of 
$^{19}$F have been restricted to explore the
role of SNe~II alone (e.g. Timmes, Woosley \& Weaver 1995; Alib\'es,
Labay \& Canal 2001).

The above problem has now been ameliorated by 
the release of the first detailed yield predictions
for fluorine production from WR and AGB stars. We are now in a position to
incorporate these yields into a Galactic chemical evolution framework, in
order to assess the respective contributions of the three putative 
fluorine production sites. To do so, we will make use of {\tt GEtool}, a
semi-analytical multi-zone
Galactic chemical evolution package which has been 
calibrated with extant observational data for the Milky Way
(\citealt{FG}; \citealt{GibsonETal}).

Specifically, in what follows, we compare the model fluorine distribution in
the Milky Way with the abundances observed by Jorissen,
Smith \& Lambert (1992) in near-solar metallicity giants. Further, our 
model predictions are contrasted with new  
fluorine determinations for
giants in the Large Magellanic Cloud (LMC) and $\omega$~Centauri 
\citep{CunhaETal}. In addition, new results for more $\omega$ Centauri giants 
from \cite{SmithETal} are included. 
The latter two systems are likely to have had {\it very}
different star formation and chemical evolution histories from those of 
the Milky Way, but despite these obvious differences, a comparison against
these new data can be valuable. In Section~2, we provide a cursory overview of
the three traditional $^{19}$F nucleosynthesis sites; the chemical evolution
code in which the nucleosynthesis products from these factories have been
implemented is described in Section~3. Our results are then presented and
summarised in Sections~4 and 5, respectively.

\section[]{Nucleosynthesis of $^{19}$F}
\subsection{Type~II Supernovae}

The massive star progenitors to SNe~II produce fluorine primarily as the result
of spallation of $^{20}$Ne by $\mu$ and $\tau$ neutrinos near the collapsed
core (\citealt{WH}; \citealt{WoosleyETal}). 
A fraction of the $^{19}$F thus created is destroyed by the subsequent
shock but most is returned to the ambient Interstellar Medium (ISM). The fluorine yields by neutrino 
spallation are very sensitive to the assumed spectra of $\mu$ and $\tau$ 
neutrinos \citep{WHW}, 
which could be nonthermal and deficient on their high-energy tails, 
lowering the equivalent temperature of the neutrinos 
in the supernova model \citep{MB}. 
An additional source of $^{19}$F derives from pre-explosive CNO 
burning in helium shell. 
However, fluorine production by neutrino spallation is largely dominant, 
as evident by comparing the models in \cite{WW}, the only ones to-date 
including neutrino process, 
and recent models which do not include neutrino nucleosynthesis 
of fluorine \citep{LC}.
Most recently, \cite{HegerETal} suggest that the relevant
neutrino cross sections need to be revised downwards; if 
confirmed, the associated SNe~II $^{19}$F yield
would decrease by $\sim$~50\%.  In light of the preliminary nature of 
the Heger et~al. claim, we retain the conservative choice offered by 
the \cite{WW} compilation.

\subsection{Asymptotic giant branch stars}

The nucleosynthesis pathways for fluorine production within AGB
stars involve both helium burning and combined hydrogen-helium burning phases 
(e.g. \citealt{ForestiniETal}; \citealt{JorissenETal}; 
\citealt{MowlaviETal})
and are companions for the nucleosynthesis by slow neutron accretion 
(s-process) 
\citep{MowlaviETal}.
Provided a suitable source of protons is available, fluorine can be synthesised
via $^{14}$N($\alpha$,$\gamma$)$^{18}$F\,($\beta^{+}$)\,
$^{18}$O(p,$\alpha$)\,$^{15}$N($\alpha$,$\gamma$)\,$^{19}$F. 
Primary sources of uncertainty in predicting fluorine nucleosynthesis 
in AGB stars relate to the adopted reaction rates, 
especially $^{14}$C$(\alpha,\gamma)^{18}$O and $^{19}$F$(\alpha,p)^{22}$Ne, 
and the treatment of the nucleosynthesis occurring during the convective 
thermal pulses. Nucleosynthesis during the interpulse periods can also 
be important if protons from the envelope are partially mixed 
in the top layers of the He intershell (partial mixing zone), as
\cite{LugaroETal} have recently demonstrated. Nucleosynthesis in this zone 
may result in a significant increase in the predicted $^{19}$F yields.
The magnitude of these 
systematic uncertainties for stellar models
with mass $\sim~3$~M$_{\odot}$ and metallicities Z = $0.004 - 0.02$ are
$\sim~50\%$, while for stellar models
with mass M = $5$~M$_{\odot}$ and metallicity Z = $0.02$ the uncertainty
is a factor of $\sim~5$, 
due to the uncertain $^{19}$F$(\alpha,p)^{22}$Ne reaction rate.
Characterising the mass- and metallicity-dependence of 
the partial mixing zone--$^{19}$F relationship needs 
to be completed before
we can assess its behaviour self-consistently within our chemical evolution
model of the Milky Way. For the present study, we have adopted the yields
presented in Appendix, based upon the Karakas \& Lattanzio (2003, and
references therein) models, which themselves do not include 
$^{19}$F nucleosynthesis via partial mixing. This choice is 
a conservative one, and thus should be considered as a lower limit to the
production of $^{19}$F from AGB stars.

For stars more massive than $\approx 4$~M$_{\odot}$, the convective
envelope is so deep that it penetrates into the top of the
hydrogen-burning shell so that nucleosynthesis actually occurs in the
envelope of the star. Such ``hot-bottom-burning'' acts to destroy $^{19}$F, 
and should be treated self-consistently within the AGB
models considered. 

\subsection{Wolf--Rayet stars}

Fluorine production in WR stars is tied to its nucleosynthesis during the 
helium-burning phase. At the end of this phase though, significant 
fluorine destruction occurs via
$^{19}$F$(\alpha,p)^{22}$Ne. Any earlier synthesised $^{19}$F
must be removed from the stellar interior in order to avoid
destruction. For massive stars to be significant contributors to net
fluorine production, they must experience mass loss on a timescale that
allows the removal of $^{19}$F before its destruction. This requirement is 
met by WR stars. 

Recently, \cite{MA} studied the role that such stars can play in the
chemical evolution of fluorine by adopting updated reaction rates coupled
with extreme mass-loss rates in not-rotating stellar models. 
They pointed out that WR mass-loss is strongly metallicity-dependent, 
and that the number of WR stars at low
metallicities is very small. Their WR yields reflect such
metallicity-dependence, with minimal fluorine returned to the ISM
at low metallicities, but significant $^{19}$F returned at solar and
super-solar metallicities. The WR yields are sensitive to the adopted reaction 
and mass-loss rates, while rotating models could favour an early entrance 
into the WR phase for a given mass, 
decrease the minimum initial mass for a star to go through a WR phase 
at a given metallicity, and open more nucleosynthetic channels 
because of the mixing induced by rotation. 
Therefore, after \cite{MA}, 
we consider the aforementioned WR yields as lower limits.

\section[]{The model}

In this study we employ {\tt GEtool}, our 
semi-analytical multi-zone chemical evolution package to model a sample
Milky Way-like disk galaxy (\citealt{FG};
\citealt{GibsonETal}). A dual--infall framework is constructed in which
the first infall episode corresponds to the formation of the halo,
and the second to the inside-out formation of the disk. 

A Kroupa, Tout \& Gilmore (1993) initial mass function (IMF) has 
been assumed, with lower and upper
mass limits of 0.08~M$_{\odot}$ and 120~M$_{\odot}$, respectively. 
Stellar yields are one of the most
important features in galactic chemical evolution models, yet
questions remain concerning the precise composition of stellar ejecta,
due to the uncertain role played by processes including mass loss,
rotation, fall-back, and the location of the mass cut, which separates
the remnant from the ejected material in SNe. The SNe~II
yields are from \cite{WW}; the yields for stars more massive than
60~M$_{\odot}$ are assumed to be mass-independent. 
Such assumption is made to avoid extreme extrapolation 
from the most massive star in the Woosley \& Weaver models (40~M$_{\odot}$) 
to the upper end of the IMF (120~M$_{\odot}$), and has 
negligible effect on the results, given the shape of the adopted IMF.

We have halved the iron yields shown in \cite{WW}, as suggested by
\cite{TimmesETal}.
The Type~Ia (SNe~Ia) yields of
\cite{IwamotoETal} were also employed. We adopted the 
metallicity-dependent yields of \cite{RV} for single stars in the mass range
$1 - 8$~M$_{\odot}$. For the purposes of this work, which focuses on fluorine, 
the choice of the Renzini \& Voli yield set does not affect the results.
Metalllicity-dependent stellar lifetimes 
have been taken from
\cite{SchallerETal}. 

We have constructed three Milky Way (MW) model variants
that differ only in their respective
treatments of $^{19}$F production:  1) MWa assumes that SNe~II are the only
source of $^{19}$F; 2) MWb includes yields from both SNe~II {\it and}
WR stars; 3) MWc includes all three sources of fluorine -
SNe~II, WR, and AGB stars. 

We end by noting that within our adopted dual--infall framework
for the chemical evolution 
of the Milky Way, our model is constrained by an array of observational
boundary conditions, including the present-day star and gas distributions
(both in density and metallicity), abundance ratio patterns, age-metallicity
relation, and G-dwarf distribution \citep{GibsonETal}.  While the modification
of any individual ingredient within model framework will have an
impact, to some degree, upon the predicted chemical evolution, this can 
only eventuate at the expense of one or more of the aforementioned 
boundary conditions that we require our model to adhere to. 
Within our framework, yield uncertainties will dominate the systematic
uncertainties for the predicted evolution of $^{19}$F.

\subsection{Fluorine yields}
\begin{figure}
\begin{center}
\includegraphics[width=0.5\textwidth]{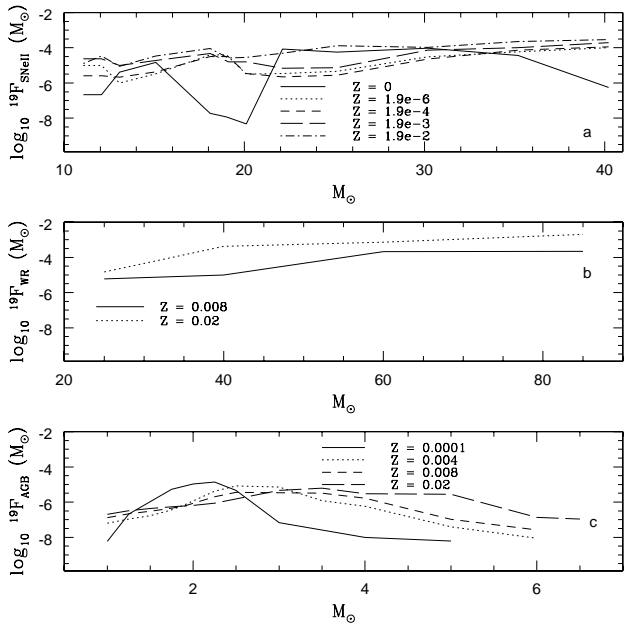}
\caption{Fluorine yields from a) SNe~II (Woosley \& Weaver 1995),
b) WR (Meynet \& Arnould 2000), and c) AGB stars (Appendix).}
\label{fig1}            
\end{center}
\end{figure}

We now summarise the $^{19}$F yields
employed in our three ``Milky Way'' models.

1) SNe~II $^{19}$F yields are taken from \cite{WW} and assumed to be 
mass-independent for stellar masses in excess of 60~M$_{\odot}$. 

2) WR $^{19}$F yields are taken 
from \cite{MA} for stellar masses in the range $25 -
120$~M$_{\odot}$: each star within this range is assumed to evolve
through the WR stage. Such simplifying assumption 
could overestimate the WR contribution to fluorine, 
even though the adopted WR yields are themselves lower limits (Section~2.3).
The WR fluorine contribution has been added to the corresponding
SNe~II contribution (which comes from a different stage of the stellar
evolution).

3) AGB $^{19}$F and oxygen yields in the $1 - 6.5$~M$_{\odot}$ mass range have
been derived from stellar models constructed with the 
Mount Stromlo Stellar Structure Code (\citealt{FL}; \citealt{KarakasETal}), 
and are presented in Appendix. 
The post-processing nucleosynthesis models 
with 74 species and time-dependent diffusive
convective mixing are described in detail in \cite{FrostETal} and \cite{KL}.

To ensure internal consistency, we have also employed the AGB 
oxygen yields \it in lieu \rm of those of Renzini \& Voli (1981),
within this mass range.

The above fluorine yields are shown in Figure~1. In Figure~2, the yields are
expressed as [F/O]\footnote{Hereafter,
[X/Y]$=log_{10}$(X/Y)$-log_{10}$(X/Y)$_{\odot}$ and 
A(X)$=12+log_{10}$(n$_{\rm X}$/n$_{\rm H}$).
An accurate determination of photospheric solar abundances 
requires detailed modeling of the solar granulation 
and accounting for departures from local thermodynamical equilibrium 
(e.g. Allende Prieto, Lambert \& Asplund 2001).
We adopt the solar fluorine abundance suggested by Cunha et~al. (2003),
and the solar iron and oxygen abundances from Holweger (2001).} 
and $\langle$[F/O]$\rangle_{IMF}$, the latter corresponding to
the mean [F/O] yields for
SNe~II and AGB stars, weighted by the IMF over the SNe~II and AGB mass
range, respectively. We have not shown a comparable entry for the WR stars as
a self-consistent treatment of the oxygen production was not included in Meynet
\& Arnould (2000). Here, oxygen has been used as the normalisation 
to make easier the comparison with the observations, 
especially in $\omega$ Centauri, though oxygen can be synthesised in various 
stellar sites, and its yields can be affected by different reaction rates 
and modeling of helium cores, semi-convection, convective boundary layers,
and mass-loss (e.g. \citealt{WHW}; \citealt{DrayETal}). 

\begin{figure}
\begin{center}
\includegraphics[width=0.5\textwidth]{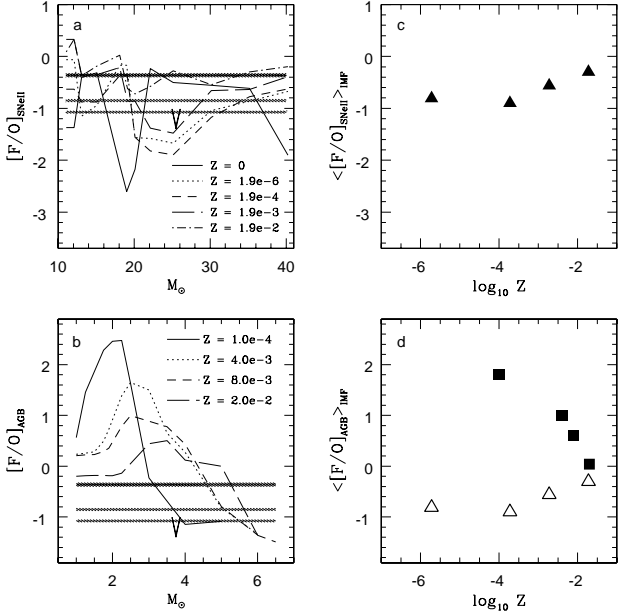}
\caption{[F/O] and $\langle$[F/O]$\rangle_{IMF}$ for SNe~II and AGB yields 
(upper and lower panels, respectively). Here, A($^{19}$F)$_{\odot} = 
4.55$ (see discussion in Cunha et al. 2003) and A(O)$_{\odot} = 8.736$ (e.g. Holweger 2001). 
The shaded regions in Figures~2a and 2b show the observed [F/O] 
in $\omega$~Cen giants (Cunha et al. 2003). The $\langle$[F/O]$\rangle_{IMF}$ 
are weighted by the IMF over the SNe~II ($11 - 40$M$_{\odot}$) and 
AGB ($1 - 6.5$~M$_{\odot}$) mass range, respectively. In Figure~2d,
both $\langle$[F/O]$_{AGB}\rangle_{IMF}$ and 
$\langle$[F/O]$_{SNe~II}\rangle_{IMF}$ are shown (closed boxes and open 
triangles, respectively).}
\label{fig2}            
\end{center}
\end{figure}

\section{Results}
\begin{figure}
\begin{center}
\includegraphics[width=0.5\textwidth]{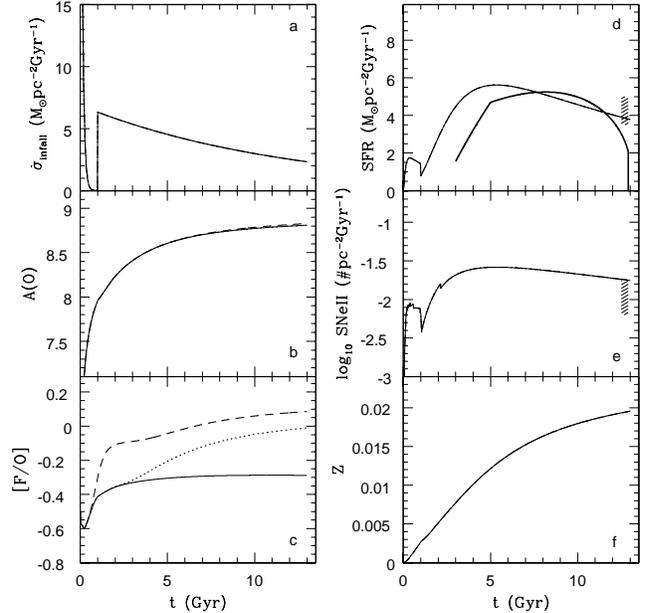}
\caption{Predicted evolution in the solar neighbourhood of a) the gas infall
rate $\dot{\sigma}_{infall}$, b) A(O), c) [F/O], d) star formation rate (SFR), 
e) SNe~II rate, and f) metallicity Z (MWa, solid line; MWb, dotted; MWc,
short-dashed). The SFR history at the solar neighbourhood obtained by 
Bertelli \& Nasi (2001) is also shown as a thick solid line in panel `d',
while the shaded region shows the range of values suggested by Rana (1991). 
A range of values corresponding to the estimated
SNe~II rate is shown in panel `e' (Cappellaro et~al. 1999).
}
\label{fig3}            
\end{center}
\end{figure}

In Figure~3, the evolution of [F/O], A(O), the gas infall rate
$\dot{\sigma}_{infall}$, the star formation rate (SFR), the SNe~II rate 
and the gas-phase global metallicity $Z$
of the three models at the solar neighbourhood are summarised. 
The empirical SFR history derived by \cite{BN} is shown as a thick solid
line in Figure~3d, while the shaded region corresponds to the range of values
suggested by \cite{Rana}. A conservative range of estimated SNe~II rates
is also shown in Figure~3e (Cappellaro, Evans \& Turatto 1999).\footnote{The
range of values shown in Figure~3e is derived from the sample
of S0a -- Sb galaxies in Cappellaro et al. (1999) - $0.42\pm0.19$~SNu, where
$1 $~SNu$=1$~SN($100$~yr)$^{-1}$($10^{10}$~L$_{\odot}^{B}$)$^{-1}$), assuming
L$_{MW}^{B}=2\times10^{10}$~L$_{\odot}^{B}$ and a galactic radial extent of
15~kpc. Given these assumptions, the estimated SNe~II
rate at the solar neighbourhood is necessarily uncertain.}. 
Figure~4 then shows the
the evolution of [F/O] versus A(O) (\it panel a\rm), and the
evolution of [F/O] versus [O/Fe] (\it panel b\rm), compared against
the IMF-weighted SNe~II yields (recall Figure~2). 

The MWa model provides a satisfactory reproduction of the estimated star
formation history and SNe~II rate in
the solar neighbourhood (Figures~3d and 3e). This model, whose only
fluorine source is SNe~II, underproduces fluorine with respect to the
abundances measured in K and M Milky Way giants observed by \cite{JorissenETal} 
and reanalysed by \cite{CunhaETal} (Fig.~4a). Fig.~4a does not show the s-process 
enriched AGB stars of spectral types MS, S, or C in \cite{JorissenETal}, 
where freshly synthesised fluorine could be mixed to the stellar surface. 
Such inclusion of self-polluted $^{19}$F-rich stars could obscure any metallicity trend.
The results of the MWa and MWb models show that the
additional contribution from WR stars 
increases [F/O] by up to factor of 2 by the present-day, but 
it is negligible
in excess of $\sim$~9~Gyr ago (Figure~3c).

The addition of both
WR {\it and} AGB sources within the MWc model 
leads to a present-day [F/O] that is
$\sim$~0.4~dex greater than in the MWa case. Further, and perhaps more
important, AGB stars are now shown to deliver
significant amounts of fluorine to the ISM during the early epochs of the
Milky Way's evolution. Such a result is entirely consistent (and expected)
given the metallicity-dependence of the AGB yields; said yields possess
[F/O] ratios which are greater at lower metallicities (recall Figure~2). 
We can conclude that it is only the addition of {\it both} the WR and AGB
contributions which allow for a significant improvement in the 
comparison between galactic models incorporating fluorine evolution and
the observational data.

\begin{figure}
\begin{center}
\includegraphics[width=0.5\textwidth]{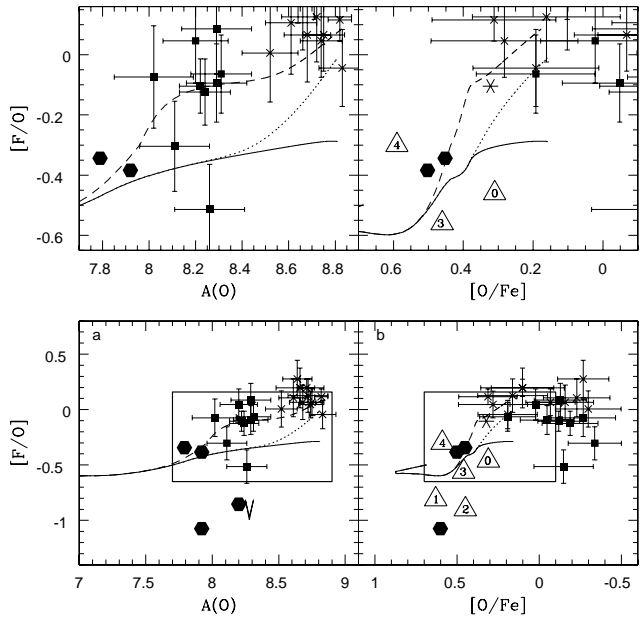}
\caption{
\it (a): \rm [F/O] as a function of A(O) for the MW models (MWa, 
solid line; MWb, dotted; MWc, short-dashed). Also shown are the
values observed in Milky Way, LMC, and $\omega$~Cen giants (crosses,
boxes and hexagons, respectively).
\it (b): \rm [F/O] as a function of [O/Fe],
compared with the IMF-weighted [O/Fe] yields for SNe~II
(open triangles). Within the open triangles, `0' corresponds to Z=0; `1', 
to Z=1.9$\times$10$^{-6}$; `2', to Z=1.9$\times$10$^{-4}$;
`3', to Z=1.9$\times$10$^{-3}$; `4', to Z=1.9$\times$10$^{-2}$.
The upper panels represent enlargements of the framed regions delineated in
the corresponding bottom panels.
}
\label{fig4}            
\end{center}
\end{figure}

\section{Discussion}

We have studied the Galactic chemical evolution of fluorine, for the first
time using new grids of stellar models which provide self-consistent
predictions of fluorine nucleosynthesis for stars in both the WR and AGB
phases of stellar evolution. We have shown that the WR contribution is 
significant at solar and super-solar metallicities because of the adopted
metallicity-dependent mass-loss prescription employed in the stellar models.
In contrast, the contribution of AGB stars to fluorine production
peaks during the early epochs of the Galaxy's evolution (again due to the
metallicity-dependent behaviour of the AGB models). In combination,
the addition of the WR and AGB yields 
leads to a significant improvement in the galactic chemical evolution models
when compared against observations. 

The comparison between our MW models and the fluorine abundances in LMC and
$\omega$~Cen giants \citep{CunhaETal} is not straightforward, as the
latter two have star formation (and therefore chemical evolution)
histories different from that of the MW. However, it is interesting to 
speculate on the possible origin of
fluorine in $\omega$~Cen, given the unique nature of this ``globular cluster''
(e.g. \citealt{S}). Specifically, $\omega$~Cen 
is the most massive Galactic cluster, and unlike most globulars, possesses a
significant spread in metallicity 
($\sim$~1.5~dex) amongst its stellar population. It has been suggested that
$\omega$~Cen is actually the remnant core of a tidally-disrupted 
dwarf galaxy \citep{BF}. Such a scenario could naturally drive radial
gas inflows to the dwarf nucleus, potentially triggering starbursts.

Interestingly, SNe~II ejecta are characterised by low
$\langle$[F/O]$\rangle_{IMF}$ (Figure~2c) and high
$\langle$[O/Fe]$\rangle_{IMF}$ (Figure~4b), whereas AGB ejecta have higher
$\langle$[F/O]$\rangle_{IMF}$ (Figure~2d). The observed $\omega$~Cen giants
have primarily low [F/O] (Figures~2a -- 2b, 4a -- 4b) and high [O/Fe]
(Figure~4b) values, consistent with a picture in which their interiors have
been polluted by the ejecta of an earlier generation of 
SNe~II, but not from a comparable generation of AGB.
Given that the observed oxygen
abundance in such $\omega$~Cen giants is similar to that seen in comparable
LMC and MW giants (Figure~4a), this would suggest that 
the chemical enrichment of $\omega$~Cen proceeded on a short timescale
(to avoid pollution from the lower mass progenitors to the AGB stars) and
in an inhomogeneous manner (given the significant scatter in observed
fluorine abundances), as discussed previously by \cite{CunhaETal}. 
Should the (downward) revised neutrino cross sections alluded
to in Section~2.1 be confirmed (Heger et~al. 2004), the concurrent 
factor of $\sim$~2 reduction in SNe~II $^{19}$F production
would improve the agreement of the model with the observed F/O ratio
in $\omega$~Cen giants.  This would consequently
strengthen our conclusions which already support a picture 
whereby these giants have been 
polluted by earlier generations of SNe~II ejecta.

\section*{Acknowledgments}
We acknowledge the financial support of the Australian Research Council 
through its Discovery Project and Linkage International schemes, and 
the Victorian Partnership for Advanced Computing through its
Expertise Grant program. We thank Maria Lugaro 
for discussion and help, 
and the referee, Marcel Arnould, for his comments. 
We also acknowledge the efforts of toothpaste 
manufacturers around the world in highlighting the importance of 
fluorine production throughout the Universe.

\bsp
\appendix
\section{Asymptotic Giant Branch $^{19}$F and $^{16,17,18}$O Yields}
The yields employed here have been derived via the following:
\begin{eqnarray}
Y_{^{19}F,~O}(Z)=& Y_{net_{^{19}F,~O}} +\nonumber\\
&X_{i_{^{19}F,~O}}(Z)\times[m_{\star} - m_{\star~rem}(Z)].
\end{eqnarray}
Here, $Y_{^{19}F,~O}(Z)$ is the overall yield, $ X_{i_{^{19}F,~O}}(Z)$ is the initial mass fraction of the element 
within a star of mass $m_{\star}$ and metallicity $Z$, 
$[m_{\star} - m_{\star~rem}(Z)]$ is the total mass ejected during the 
stellar lifetime, and $Y_{net_{^{19}F,~O}}$ is the net
yield. We calculate $X_{i_{O}}$ and $Y_{net_{O}}$ as, respectively: 
\begin{eqnarray}
X_{i_{O}} = X_{i_{^{16}O}} + X_{i_{^{17}O}} + X_{i_{^{18}O}};\\
Y_{net_{O}} = Y_{net_{^{16}O}} + Y_{net_{^{17}O}} + Y_{net_{^{18}O}}.
\end{eqnarray}
The yields are the result of full evolutionary calculations using 
the Mount Stromlo Stellar Structure Code 
(e.g. \citealt{KL}). We use the standard Reimers mass-loss formula on the 
first giant branch and the \cite{VW} formula during 
the AGB evolution. Opacities are from OPAL \citep{IR}.
The models with Z~=~0.02 and 0.0001 used scaled solar abundances, whereas 
those for Z~=~0.004 and 0.008 are appropriate to the Small and Large 
Magellanic Clouds, respectively, and are taken from \cite{RD}. 
Numerical problems during the third dredge-up are handled in the 
way described in \cite{FL}. A mixing length 
of 1.75 pressure scale-heights has been used. 
A main uncertainty in the predicted 
yields for fluorine is the occurrence and dimension of the partial mixing zone. 
Note that this partial mixing zone 
was ignored in the models presented here. 
Primary sources of uncertainty in predicting fluorine nucleosynthesis 
in AGB stars relate to the adopted reaction rates, 
especially $^{14}$C$(\alpha,\gamma)^{18}$O and $^{19}$F$(\alpha,p)^{22}$Ne, 
and the treatment of the nucleosynthesis occurring during the convective 
thermal pulses and the interpulse periods (\citealt{LugaroETal}; see also 
Section~2.2).

\begin{table*}
\begin{center}
\caption{AGB $^{19}$F and $^{16,17,18}$O yields (M$_{\odot}$).}
\begin{tabular}{@{}llllll@{}}
&&&&&\\
\hline
$Z$&&$X_{i_{^{19}F}}$&$X_{i_{^{16}O}}$&$X_{i_{^{17}O}}$&$X_{i_{^{18}O}}$\\
0.0001&&2.31800e-09&4.44786e-05&1.93800e-08&1.08000e-07\\
\hline
$m_{\star}$&$m_{\star~rem}$&$Y_{net~^{19}F_{AGB}}$&$Y_{net~^{16}O_{AGB}}$&$Y_{net~^{17}O_{AGB}}$&$Y_{net~^{18}O_{AGB}}$\\
1.0~&0.65~&~0.51379E-08&0.52869E-05&0.38546E-07&-0.59145E-08\\
1.25&0.65~&~0.21202E-06&0.69519E-04&0.72848E-07&-0.14220E-07\\
1.75&0.67~&~0.54275E-05&0.31623E-03&0.60806E-06&~0.21305E-07\\
2.0~&0.70~&~0.10527E-04&0.41221E-03&0.78911E-06&~0.66983E-07\\
2.25&0.72~&~0.13578E-04&0.51952E-03&0.41433E-06&~0.99055E-07\\
2.5~&0.737&~0.45574E-05&0.13068E-02&0.74846E-06&-0.17541E-07\\
3.0~&0.820&~0.62016E-07&0.13776E-02&0.20197E-05&-0.23173E-06\\
4.0~&0.868&~0.27367E-08&0.16630E-02&0.36273E-05&-0.33571E-06\\
5.0~&0.924&-0.34142E-08&0.76699E-03&0.19749E-05&-0.41578E-06\\
\hline
\hline
$Z$&&$X_{i_{^{19}F}}$&$X_{i_{^{16}O}}$&$X_{i_{^{17}O}}$&$X_{i_{^{18}O}}$\\
0.004~&&1.69218E-07&1.28324E-03&1.41477E-06&7.88419E-06\\
\hline
$m_{\star}$&$m_{\star~rem}$&$Y_{net~^{19}F_{AGB}}$&$Y_{net~^{16}O_{AGB}}$&$Y_{net~^{17}O_{AGB}}$&$Y_{net~^{18}O_{AGB}}$\\
1.0~&0.63~&~0.94462E-09&-0.29158E-05&0.22240E-06&-0.25334E-06\\
1.25&0.64~&~0.62277E-08&-0.20669E-05&0.49241E-06&-0.89206E-06\\
1.5~&0.646&~0.22912E-07&~0.19546E-04&0.15093E-05&-0.15635E-05\\
1.75&0.65~&~0.17284E-06&~0.75738E-04&0.44914E-05&-0.25162E-05\\
1.9~&0.65~&~0.49608E-06&~0.95973E-04&0.75796E-05&-0.32065E-05\\
2.25&0.66~&~0.38720E-05&~0.14749E-03&0.11186E-04&-0.43546E-05\\
2.5~&0.68~&~0.80892E-05&~0.93431E-04&0.98989E-05&-0.54950E-05\\
3.0~&0.73~&~0.68954E-05&~0.16497E-04&0.10204E-04&-0.69500E-05\\
3.5~&0.82~&~0.76727E-06&~0.62994E-04&0.43600E-05&-0.19736E-04\\
4.0~&0.86~&~0.55804E-07&~0.68820E-05&0.17349E-05&-0.24560E-04\\ 
5.0~&0.91~&-0.65262E-06&-0.19071E-02&0.20350E-05&-0.32029E-04\\
6.0~&0.97~&-0.84234E-06&-0.39748E-02&0.10162E-05&-0.39405E-04\\
\hline
\hline
$Z$&&$X_{i_{^{19}F}}$&$X_{i_{^{16}O}}$&$X_{i_{^{17}O}}$&$X_{i_{^{18}O}}$\\
0.008~&&3.25423E-07&2.64027E-03&2.72075E-06&1.51621E-05\\
\hline
$m_{\star}$&$m_{\star~rem}$&$Y_{net~^{19}F_{AGB}}$&$Y_{net~^{16}O_{AGB}}$&$Y_{net~^{17}O_{AGB}}$&$Y_{net~^{18}O_{AGB}}$\\
1.0~&0.5998&~0.23717E-08&-0.26083E-05&0.28610E-06&-0.39665E-06\\
1.25&0.61~~&~0.11401E-07&-0.40814E-05&0.65704E-06&-0.15151E-05\\
1.5~&0.63~~&~0.20168E-07&-0.45432E-05&0.19727E-05&-0.28219E-05\\
1.75&0.64~~&~0.90125E-07&~0.22020E-04&0.60267E-05&-0.45889E-05\\
1.9~&0.64~~&~0.18749E-06&~0.38094E-04&0.95185E-05&-0.53377E-05\\
2.25&0.65~~&~0.14941E-05&~0.85696E-05&0.16553E-04&-0.74497E-05\\
2.5~&0.67~~&~0.30357E-05&-0.12452E-03&0.14976E-04&-0.90888E-05\\
3.5~&0.77~~&~0.22972E-05&-0.37752E-03&0.16132E-04&-0.14133E-04\\
4.0~&0.84~~&~0.66878E-06&-0.30252E-03&0.14747E-04&-0.23226E-04\\
5.0~&0.89~~&-0.12324E-05&-0.24193E-02&0.68396E-05&-0.61869E-04\\
6.0~&0.95~~&-0.16173E-05&-0.57759E-02&0.80141E-05&-0.75982E-04\\
\hline
\hline
$Z$&&$X_{i_{^{19}F}}$&$X_{i_{^{16}O}}$&$X_{i_{^{17}O}}$&$X_{i_{^{18}O}}$\\
0.02~~&&4.63728E-07&9.60266E-03&3.87707E-06&2.16506E-05\\
\hline
$m_{\star}$&$m_{\star~rem}$&$Y_{net~^{19}F_{AGB}}$&$Y_{net~^{16}O_{AGB}}$&$Y_{net~^{17}O_{AGB}}$&$Y_{net~^{18}O_{AGB}}$\\
1.0~&0.57309&~0.36025E-08&-0.61933E-06&0.34219E-06&-0.41416E-06\\
1.25&0.578~~&~0.15858E-07&-0.22254E-05&0.97603E-06&-0.18066E-05\\
1.5~&0.60~~~&~0.25014E-07&-0.32904E-05&0.31910E-05&-0.33975E-05\\
1.75&0.63~~~&~0.29957E-07&-0.54827E-05&0.90611E-05&-0.52622E-05\\
1.9~&0.636~~&~0.28199E-07&-0.10482E-04&0.14418E-04&-0.60505E-05\\
2.0~&0.64~~~&~0.27049E-07&-0.17893E-04&0.19332E-04&-0.66536E-05\\
2.25&0.65~~~&~0.12064E-06&-0.20851E-03&0.44894E-04&-0.81507E-05\\
3.0~&0.682~~&~0.34981E-05&-0.11882E-02&0.51877E-04&-0.13306E-04\\
3.5~&0.716~~&~0.49320E-05&-0.17086E-02&0.57427E-04&-0.17219E-04\\
4.0~&0.791~~&~0.14844E-05&-0.16979E-02&0.55128E-04&-0.19531E-04\\
5.0~&0.878~~&~0.86376E-06&-0.39096E-02&0.46522E-04&-0.70118E-04\\
6.0~&0.929~~&-0.22147E-05&-0.83059E-02&0.78725E-04&-0.10882E-03\\
6.5~&0.964~~&-0.24583E-05&-0.97865E-02&0.98794E-04&-0.11879E-03\\
\hline
\end{tabular}
\end{center}
\end{table*}

\label{lastpage}

\end{document}